%Paper: hep-th/9210142
%From: Albrecht Klemm <klemm@string3.harvard.edu>
%Date: Tue, 27 Oct 92 23:39:31 -0400

%%%%%%%%%%%%%%%%%%  tex macros for preprints, cm version %%%%%%%%%%%%%%%
%                      P. Ginsparg
%                      last updated 7/89
%                      nonharvard users say "b" in response
%                      to query, and only plain.tex is needed
%
\newbox\leftpage \newdimen\fullhsize \newdimen\hstitle \newdimen\hsbody
\tolerance=1000\hfuzz=2pt
\def\printertype{ps: }% default postscript
\def\qms{\def\printertype{qms: }% qms lasergrafix
\ifx\answ\bigans\else\voffset=-.4truein\hoffset=.125truein\fi}
\def\us#1{{\underline{#1}}}
\def\bigans{b }
%\message{ big or little (b/l)? }\read-1 to\answ
%
\def\answ{b }
%
%\ifx\answ\bigans\message{(This will come out unreduced.}
%\magnification=1200\baselineskip=16pt plus 2pt minus 1pt
%       Unpleasantness in calling in abstract and title fonts
\font\titlerm=cmr10 scaled\magstep3 \font\titlerms=cmr7 scaled\magstep3
\font\titlermss=cmr5 scaled\magstep3 \font\titlei=cmmi10 scaled\magstep3
\font\titleis=cmmi7 scaled\magstep3 \font\titleiss=cmmi5 scaled\magstep3
\font\titlesy=cmsy10 scaled\magstep3 \font\titlesys=cmsy7 scaled\magstep3
\font\titlesyss=cmsy5 scaled\magstep3 \font\titleit=cmti10 scaled\magstep3
\skewchar\titlei='177 \skewchar\titleis='177 \skewchar\titleiss='177
\skewchar\titlesy='60 \skewchar\titlesys='60 \skewchar\titlesyss='60
\def\titlefont{\def\rm{\fam0\titlerm}% switch to title font
\textfont0=\titlerm \scriptfont0=\titlerms \scriptscriptfont0=\titlermss
\textfont1=\titlei \scriptfont1=\titleis \scriptscriptfont1=\titleiss
\textfont2=\titlesy \scriptfont2=\titlesys \scriptscriptfont2=\titlesyss
\textfont\itfam=\titleit \def\it{\fam\itfam\titleit} \rm}
\font\absrm=cmr10 scaled\magstep1 \font\absrms=cmr7 scaled\magstep1
\font\absrmss=cmr5 scaled\magstep1 \font\absi=cmmi10 scaled\magstep1
\font\absis=cmmi7 scaled\magstep1 \font\absiss=cmmi5 scaled\magstep1
\font\abssy=cmsy10 scaled\magstep1 \font\abssys=cmsy7 scaled\magstep1
\font\abssyss=cmsy5 scaled\magstep1 \font\absbf=cmbx10 scaled\magstep1
\skewchar\absi='177 \skewchar\absis='177 \skewchar\absiss='177
\skewchar\abssy='60 \skewchar\abssys='60 \skewchar\abssyss='60
\font\bigit=cmti10 scaled \magstep1
\def\abstractfont{\def\rm{\fam0\absrm}% switch to abstract font
\textfont0=\absrm \scriptfont0=\absrms \scriptscriptfont0=\absrmss
\textfont1=\absi \scriptfont1=\absis \scriptscriptfont1=\absiss
\textfont2=\abssy \scriptfont2=\abssys \scriptscriptfont2=\abssyss
\textfont\itfam=\bigit \def\it{\fam\itfam\bigit}
\textfont\bffam=\absbf \def\bf{\fam\bffam\absbf} \rm} %\fi
%

%
%---------------------------------------------------------------------
%
\abstractfont
\baselineskip=20pt plus 2pt minus 1pt
\hsbody=\hsize \hstitle=\hsize %take default values for unreduced format
%
%\else\message{(This will be reduced.} \let\lr=L
%\magnification=1000\baselineskip=16pt plus 2pt minus 1pt
%\voffset=-.31truein\vsize=7truein\hoffset=-.59truein% apple lw
%\hstitle=8truein\hsbody=4.75truein\fullhsize=10truein\hsize=\hsbody
%%
%\output={\ifnum\pageno=0 %%% This is the HUTP version
%  \shipout\vbox{\special{\printertype landscape}\makeheadline
%    \hbox to \fullhsize{\hfill\pagebody\hfill}}\advancepageno
%  \else
%  \almostshipout{\leftline{\vbox{\pagebody\makefootline}}}\advancepageno
%  \fi}
%\def\almostshipout#1{\if L\lr \count1=1
\message{[\the\count0.\the\count1]}
%      \global\setbox\leftpage=#1 \global\let\lr=R
%  \else \count1=2
%    \shipout\vbox{\special{\printertype landscape}
%      \hbox to\fullhsize{\box\leftpage\hfil#1}}  \global\let\lr=L\fi}
%\fi
%
%---------------------------------------------------------------------
\catcode`\@=11 % This allows us to modify PLAIN macros.
\newcount\yearltd\yearltd=\year\advance\yearltd by -1900

%
% restores pagenumbers
%
%       use following instead of \Date on the preliminary draft,
%       puts date/time on each page in big mode, writes labels in margins

\def\draftmode{\message{ DRAFTMODE }\def\draftdate{{\rm preliminary draft:
\number\month/\number\day/\number\yearltd\ \ \hourmin}}%
\headline={\hfil\draftdate}\writelabels\baselineskip=20pt plus 2pt minus
2pt
 {\count255=\time\divide\count255 by 60 \xdef\hourmin{\number\count255}
  \multiply\count255 by-60\advance\count255 by\time
  \xdef\hourmin{\hourmin:\ifnum\count255<10 0\fi\the\count255}}}
%       use \nolabels to get rid of eqn, ref, and fig labels in draft mode
\def\nolabels{\def\wrlabel##1{}\def\eqlabel##1{}\def\reflabel##1{}}
\def\writelabels{\def\wrlabel##1{\leavevmode\vadjust
{\rlap{\smash{\line{{\escapechar=`
\hfill\rlap{\hskip.03in\string##1}}}}}}}%
\def\eqlabel##1{{\escapechar=` \rlap{\hskip.09in\string##1}}}%
\def\reflabel##1{\noexpand\llap{\string\string\string##1}}}
\nolabels
%
% tagged sec numbers
\global\newcount\secno \global\secno=0
\global\newcount\meqno \global\meqno=1
\def\newsec#1{\global\advance\secno by1\message{(\the\secno. #1)}
\global\subsecno=0\xdef\secsym{\the\secno.}\global\meqno=1
%\ifx\answ\bigans \vfill\eject \else \bigbreak\bigskip \fi  %if desired
\bigbreak\bigskip\noindent{\bf\the\secno. #1}\writetoca{{\secsym} {#1}}
\par\nobreak\medskip\nobreak}
\xdef\secsym{}
\global\newcount\subsecno \global\subsecno=0
\def\subsec#1{\global\advance\subsecno by1\message{(\secsym\the\subsecno.
#1)}
\bigbreak\noindent{\it\secsym\the\subsecno. #1}\writetoca{\string\quad
{\secsym\the\subsecno.} {#1}}\par\nobreak\medskip\nobreak}
\def\appendix#1#2{\global\meqno=1\global\subsecno=0\xdef\secsym{\hbox{#1.}}
\bigbreak\bigskip\noindent{\bf Appendix #1. #2}\message{(#1. #2)}
\writetoca{Appendix {#1.} {#2}}\par\nobreak\medskip\nobreak}
%
%       \eqn\label{a+b=c}gives displayed equation, numbered
%consecutively within sections.
%     \eqnn and \eqna define labels in advance (of eqalign?)
%
\def\eqnn#1{\xdef #1{(\secsym\the\meqno)}\writedef{#1\leftbracket#1}%
\global\advance\meqno by1\wrlabel#1}
\def\eqna#1{\xdef #1##1{\hbox{$(\secsym\the\meqno##1)$}}
\writedef{#1\numbersign1\leftbracket#1{\numbersign1}}%
\global\advance\meqno by1\wrlabel{#1$\{\}$}}
\def\eqn#1#2{\xdef #1{(\secsym\the\meqno)}\writedef{#1\leftbracket#1}%
\global\advance\meqno by1$$#2\eqno#1\eqlabel#1$$}
%
% footnotes
\newskip\footskip\footskip14pt plus 1pt minus 1pt %sets footnote
baselineskip
\def\f@@t{\baselineskip\footskip\bgroup\aftergroup\@foot\let\next}
\setbox\strutbox=\hbox{\vrule height9.5pt depth4.5pt width0pt}
\global\newcount\ftno \global\ftno=0
\def\foot{\global\advance\ftno by1\footnote{$^{\the\ftno}$}}
%
%say \footend to put footnotes at end
%will cause problems if \ref used inside \foot, instead use \nref before
\newwrite\ftfile
\def\footend{\def\foot{\global\advance\ftno by1\chardef\wfile=\ftfile
$^{\the\ftno}$\ifnum\ftno=1\immediate\openout\ftfile=foots.tmp\fi%
\immediate\write\ftfile{\noexpand\smallskip%
\noexpand\item{f\the\ftno:\ }\pctsign}\findarg}%
\def\footatend{\vfill\eject\immediate\closeout\ftfile{\parindent=20pt
\centerline{\bf Footnotes}\nobreak\bigskip\input foots.tmp }}}
\def\footatend{}
%
%     \ref\label{text}
% generates a number, assigns it to \label, generates an entry.
% To list the refs on a separate page,  \listrefs
%
\global\newcount\refno \global\refno=1
\newwrite\rfile
\def\ref{[\the\refno]\nref}
\def\nref#1{\xdef#1{[\the\refno]}\writedef{#1\leftbracket#1}%
\ifnum\refno=1\immediate\openout\rfile=refs.tmp\fi
\global\advance\refno by1\chardef\wfile=\rfile\immediate
\write\rfile{\noexpand\item{#1\ }\reflabel{#1\hskip.31in}\pctsign}\findarg}
%horrible hack to sidestep tex \write limitation
\def\findarg#1#{\begingroup\obeylines\newlinechar=`\^^M\pass@rg}
{\obeylines\gdef\pass@rg#1{\writ@line\relax #1^^M\hbox{}^^M}%
\gdef\writ@line#1^^M{\expandafter\toks0\expandafter{\striprel@x #1}%
\edef\next{\the\toks0}\ifx\next\em@rk\let\next=\endgroup\else\ifx\next\empty%
\else\immediate\write\wfile{\the\toks0}\fi\let\next=\writ@line\fi\next\relax}}
\def\striprel@x#1{} \def\em@rk{\hbox{}}

\def\addref#1{\immediate\write\rfile{\noexpand\item{}#1}} %now unnecessary
\def\startrefs#1{\immediate\openout\rfile=refs.tmp\refno=#1}
\def\xref{\expandafter\xr@f}\def\xr@f[#1]{#1}
\def\refs#1{[\r@fs #1{\hbox{}}]}
\def\r@fs#1{\edef\next{#1}\ifx\next\em@rk\def\next{}\else
\ifx\next#1\xref #1\else#1\fi\let\next=\r@fs\fi\next}
%

%
% this is ugly, but moore insists
\newwrite\ffile\global\newcount\figno \global\figno=1
\def\fig{fig.~\the\figno\nfig}
\def\nfig#1{\xdef#1{fig.~\the\figno}%
\writedef{#1\leftbracket fig.\noexpand~\the\figno}%
\ifnum\figno=1\immediate\openout\ffile=figs.tmp\fi\chardef\wfile=\ffile%
\immediate\write\ffile{\noexpand\medskip\noexpand\item{Fig.\ \the\figno. }
\reflabel{#1\hskip.55in}\pctsign}\global\advance\figno by1\findarg}
\def\xfig{\expandafter\xf@g}\def\xf@g fig.\penalty\@M\ {}
\def\figs#1{figs.~\f@gs #1{\hbox{}}}
\def\f@gs#1{\edef\next{#1}\ifx\next\em@rk\def\next{}\else
\ifx\next#1\xfig #1\else#1\fi\let\next=\f@gs\fi\next}
\newwrite\lfile
{\escapechar-1\xdef\pctsign{\string\%}\xdef\leftbracket{\string\{}
\xdef\rightbracket{\string\}}\xdef\numbersign{\string\#}}

\def\writestop{\def\writestoppt{\immediate\write\lfile{\string\pageno%
\the\pageno\string\startrefs\leftbracket\the\refno\rightbracket%
\string\def\string\secsym\leftbracket\secsym\rightbracket%
\string\secno\the\secno\string\meqno\the\meqno}\immediate\closeout\lfile}}
\def\writestoppt{}\def\writedef#1{}
\def\seclab#1{\xdef
#1{\the\secno}\writedef{#1\leftbracket#1}\wrlabel{#1=#1}}
\def\subseclab#1{\xdef #1{\secsym\the\subsecno}%
\writedef{#1\leftbracket#1}\wrlabel{#1=#1}}
\newwrite\tfile \def\writetoca#1{}
\def\leaderfill{\leaders\hbox to 1em{\hss.\hss}\hfill}
%use this to write file with table of contents
\def\writetoc{\immediate\openout\tfile=toc.tmp
   \def\writetoca##1{{\edef\next{\write\tfile{\noindent ##1
   \string\leaderfill {\noexpand\number\pageno} \par}}\next}}}
%       and this lists table of contents on second pass

%
\catcode`\@=12 % at signs are no longer letters
%
%Unpleasantness in calling in abstract and title fonts
\ifx\answ\bigans
\font\titlerm=cmr10 scaled\magstep3 \font\titlerms=cmr7 scaled\magstep3
\font\titlermss=cmr5 scaled\magstep3 \font\titlei=cmmi10 scaled\magstep3
\font\titleis=cmmi7 scaled\magstep3 \font\titleiss=cmmi5 scaled\magstep3
\font\titlesy=cmsy10 scaled\magstep3 \font\titlesys=cmsy7 scaled\magstep3
\font\titlesyss=cmsy5 scaled\magstep3 \font\titleit=cmti10 scaled\magstep3
\else
\font\titlerm=cmr10 scaled\magstep4 \font\titlerms=cmr7 scaled\magstep4
\font\titlermss=cmr5 scaled\magstep4 \font\titlei=cmmi10 scaled\magstep4
\font\titleis=cmmi7 scaled\magstep4 \font\titleiss=cmmi5 scaled\magstep4
\font\titlesy=cmsy10 scaled\magstep4 \font\titlesys=cmsy7 scaled\magstep4
\font\titlesyss=cmsy5 scaled\magstep4 \font\titleit=cmti10 scaled\magstep4
\font\absrm=cmr10 scaled\magstep1 \font\absrms=cmr7 scaled\magstep1
\font\absrmss=cmr5 scaled\magstep1 \font\absi=cmmi10 scaled\magstep1
\font\absis=cmmi7 scaled\magstep1 \font\absiss=cmmi5 scaled\magstep1
\font\abssy=cmsy10 scaled\magstep1 \font\abssys=cmsy7 scaled\magstep1
\font\abssyss=cmsy5 scaled\magstep1 \font\absbf=cmbx10 scaled\magstep1
\skewchar\absi='177 \skewchar\absis='177 \skewchar\absiss='177
\skewchar\abssy='60 \skewchar\abssys='60 \skewchar\abssyss='60
\fi
\skewchar\titlei='177 \skewchar\titleis='177 \skewchar\titleiss='177
\skewchar\titlesy='60 \skewchar\titlesys='60 \skewchar\titlesyss='60
\def\titlefont{\def\rm{\fam0\titlerm}% switch to title font
\textfont0=\titlerm \scriptfont0=\titlerms \scriptscriptfont0=\titlermss
\textfont1=\titlei \scriptfont1=\titleis \scriptscriptfont1=\titleiss
\textfont2=\titlesy \scriptfont2=\titlesys \scriptscriptfont2=\titlesyss
\textfont\itfam=\titleit \def\it{\fam\itfam\titleit} \rm}
%
%\ifx\answ\bigans\def\abstractfont{\tenpoint}\else
\def\abstractfont{\def\rm{\fam0\absrm}% switch to abstract font
\textfont0=\absrm \scriptfont0=\absrms \scriptscriptfont0=\absrmss
\textfont1=\absi \scriptfont1=\absis \scriptscriptfont1=\absiss
\textfont2=\abssy \scriptfont2=\abssys \scriptscriptfont2=\abssyss
\textfont\itfam=\bigit \def\it{\fam\itfam\bigit}
\textfont\bffam=\absbf \def\bf{\fam\bffam\absbf} \rm}
%\fi
%

%
%---------------------------------------------------------------------
%

\hyphenation{anom-aly anom-alies coun-ter-term coun-ter-terms}
\def\inv{^{\raise.15ex\hbox{${\scriptscriptstyle -}$}\kern-.40em 1}}

\def\Dsl{\,\raise.15ex\hbox{/}\mkern-13.5mu D} %this one can be subscripted
\def\dsl{\raise.15ex\hbox{/}\kern-.57em\partial}

\font\bigit=cmti10 scaled \magstep1
 %pound sterling
\def\lspace{\ifx\answ\bigans{}\else\qquad\fi}
\def\lbspace{\ifx\answ\bigans{}\else\hskip-.2in\fi} % $$\lbspace...$$
\def\boxeqn#1{\vcenter{\vbox{\hrule\hbox{\vrule\kern3pt\vbox{\kern3pt
\hbox{${\displaystyle #1}$}\kern3pt}\kern3pt\vrule}\hrule}}}
\def\mbox#1#2{\vcenter{\hrule \hbox{\vrule height#2in
\kern#1in \vrule} \hrule}}    %e.g. \mbox{.1}{.1}
%matters of taste
%\def\tilde{\widetilde} \def\bar{\overline} \def\hat{\widehat}
%
% some sample definitions
  %     curly letters

\def\darr#1{\raise1.5ex\hbox{$\leftrightarrow$}\mkern-16.5mu #1}
 %pound sterling

 %puts a small half in a displayed eqn
\def\roughly#1{\raise.3ex\hbox{$#1$\kern-.75em\lower1ex\hbox{$\sim$}}}

\def\frac#1#2{{#1\over#2}}

\def\journal#1&#2(#3){\unskip, #1~\bf #2 \rm(19#3) }
\def\andjournal#1&#2(#3){\sl #1~\bf #2 \rm (19#3) }

\def\exp{{\rm exp}}

\catcode`\@=11
\def\slash#1{\mathord{\mathpalette\c@ncel{#1}}}
\overfullrule=0pt
\def\steepslash{\c@ncel}
\def\frac#1#2{{#1\over #2}}

\def\p {\partial}
\def\Z{{\bf Z}}

\def\N{{\bf N}}
\def\R{{\bf R}}
\def\inbar{\,\vrule height1.5ex width.4pt depth0pt}
\def\IB{\relax{\rm I\kern-.18em B}}
\def\IC{\relax\hbox{$\inbar\kern-.3em{\rm C}$}}
\def\IP{\relax{\rm I\kern-.18em P}}
\def\IR{\relax{\rm I\kern-.18em R}}
\def\IZ{\relax\ifmmode\mathchoice
{\hbox{Z\kern-.4em Z}}{\hbox{Z\kern-.4em Z}}
{\lower.9pt\hbox{Z\kern-.4em Z}}
{\lower1.2pt\hbox{Z\kern-.4em Z}}\else{Z\kern-.4em Z}\fi}

\catcode`\@=12

%                      Zeitschriften:
\def\npb#1(#2)#3{{ Nucl. Phys. }{\bf B#1} (#2) #3}
\def\plb#1(#2)#3{{ Phys. Lett. }{\bf #1B} (#2) #3}
\def\pla#1(#2)#3{{ Phys. Lett. }{\bf #1A} (#2) #3}
\def\prl#1(#2)#3{{ Phys. Rev. Lett. }{\bf #1} (#2) #3}
\def\mpla#1(#2)#3{{ Mod. Phys. Lett. }{\bf A#1} (#2) #3}
\def\ijmpa#1(#2)#3{{ Int. J. Mod. Phys. }{\bf A#1} (#2) #3}
\def\cmp#1(#2)#3{{ Comm. Math. Phys. }{\bf #1} (#2) #3}
\def\cqg#1(#2)#3{{ Class. Quantum Grav. }{\bf #1} (#2) #3}
\def\jmp#1(#2)#3{{ J. Math. Phys. }{\bf #1} (#2) #3}
\def\anp#1(#2)#3{{ Ann. Phys. }{\bf #1} (#2) #3}

\def\a{\alpha}

\def\g{\gamma}
\def\o{\omega}

\def\p1{\phi_1}
\def\p2{\phi_2}

\def\({\lbrack}
\def\){\rbrack}
\def\mat{\matrix}

\footline{\hss\tenrm--\folio\--\hss}

\centerline{\titlefont
Recent Efforts in the Computation of String Couplings\footnote{$^*$}
{\tenrm Partially supported by the Deutsche Forschungsgemeinschaft.}
\footnote{$^\dagger$}
{\tenrm Contribution to the `International Conference on Modern
Problems
in Quantum Field Theorie, Strings and Quantum Gravity', Kiev,
June 8-17,1992.}}
\vskip1cm

\centerline{Albrecht Klemm
and Stefan Theisen}
\medskip\centerline{Sektion Physik der Universit\"at M\"unchen}
\centerline{Theresienstra\ss e 37, D - 8000 M\"unchen 2, FRG}
\vskip .5in

{\noindent {\bf Abstract:} We review recent advances towards the
computation of string couplings.
Duality symmetry, mirror symmetry, Picard-Fuchs equations, etc.
are some of the tools.}
\vskip1.5cm

One of the main topics of this conference was the matrix
model approach to non-critical strings. There the outstanding
open problem is to go above $c=1$. Here we want to review
some recent progress in the `old-fashioned' formulation of
critical string theory with $(c,\bar c)=(15,26)$ (in the
case of the heterotic string).  Since the description of the
space-time degrees of freedom only uses up $(6,4)$ units
of the central charge, one uses the remaining $(9,22)$
to describe internal degrees of freedom (gauge
symmetries). We will not discuss any of the
conditions which have to be imposed on the string vacua,
such as absence of tachyons, modular invariance, etc.
In the class of models we will mainly be concerned with,
namely Calabi-Yau compactifications \ref\huebsch{For a review with
an extensive list of references, consult T. H\"ubsch,
{\it Calabi-Yau Manifolds}, World Scientific, 1991.},
they are all satisfied.
We will rather address the problem of how to close the gap
between the formal description and classification
of string vacua and their possible role in a realistic
description of particle physics. Even if one finds a model
with the desired particle content and gauge symmetry, one is
still confronted with the problem of computing the couplings,
which determine masses, mixing angles, patterns of symmetry
breaking etc. These couplings will depend on the moduli of
the string model, which, in the conformal field theory language,
correspond to the exactly marginal operators, or, in the
Calabi-Yau context, to the harmonic (1,1) and (2,1) forms, which
describe deformations of the K\"ahler class and the
complex structure, respectively.
Indeed, if one varies the metric $g_{i\bar\jmath}$
($i,\bar\jmath=1,2,3$)
on the Calabi-Yau space, preserving Ricci flatness, one
finds that $i\delta g_{i\bar\jmath}$ (corresponding to variations
of the K\"ahler class) are (real) components of harmonic (1,1)
forms,
whereas $\Omega_{ij}{}^{\bar l}\delta g_{\bar l\bar k}$
(corresponding to variations of the complex structure)
are (complex) components of harmonic (2,1) forms.
Here $\Omega_{ijk}=g_{k\bar l}\Omega_{ij}{}^{\bar l}$ is the
unique (up to a scale) covariantly constant three form which
is always present. Recall that $h_{3,0}=1$ and
$h_{1,0}=h_{2,0}=0$ on Ricci flat Calabi-Yau three-folds.
($h_{i,j}$ denotes the number of harmonic $(i,j)$ forms.)
{}From its equation of motion one finds that the internal components
of the anti-symmetric tensor field also have to correspond to
harmonic forms. Since there are no harmonic $(2,0)$ forms,
we can take the mixed components $B_{i\bar j}$ to complexify
the components of the harmonic (1,1) forms.
In (2,2) compactifications
which are the ones which have been most intensively studied
to date, the two types of moduli are related by world sheet
supersymmetry to the matter fields, which transform as
$\overline{27}$ and $27$ of $E_6$.
In the conformal field theory language the moduli correspond to
truely marginal operators.
In a low energy effective field
theory description, which includes all the light states, but
having integrated out all heavy ($>m_{\rm Planck}$) string modes,
the moduli appear as massless neutral scalar fields with
perturbatively vanishing potential. Thus, the
strength of the couplings, such
as the Yukawa couplings, which do depend on the moduli, are
undetermined. Only if the vacuum expectation value of the
moduli fields is fixed by a non-perturbative potential do the
couplings take fixed values, which could then be compared
with experiment.
To get the physical couplings one also needs to
determine the K\"ahler metric for all the fields
involved in order to normalize them properly.

Generic string models are believed to possess duality
symmetry\ref\dualgen{K. Kikkawa and M. Yamasaki, \plb149(1984)357;
N. Sakai and I. Senda, Prog. Theor. Phys. {\bf 75} (1984) 692;
V. P. Nair, A. Shapere, A. Strominger and F. Wilczek,
\npb287(1987)402; A. Giveon, E. Rabinovici and G. Veneziano,
\npb322(1989)167.},
which is a discrete symmetry on moduli space
that leaves the spectrum as well as the
interactions invariant and whose origin is tied to the fact that
strings
are one-dimensional extended objects. This symmetry has been
explicitly
found in simple models, such as the compactification on tori and
their orbifolds
\ref\orbdual{  M. Dine, P. Huet and N. Seiberg, \npb322(1989)301;
J.Lauer, J. Mas and H. P. Nilles, \plb226(1989)251;
M. Spali\'nski, \plb275(1992)47.},
but more recently for some simple Calabi-Yau
compactifications
\ref\cdgp{P. Candelas, X. de la Ossa, P. Green and
L. Parkes, \npb359(1991)21.}
\ref\kt{A. Klemm and S. Theisen, {\it Considerations of One-Modulus
Calabi-Yau Compactifications: Picard-Fuchs equations, K\"ahler
Potentials and Mirror Maps}, Nucl. Phys. B (in press).}
\ref\afont{A. Font, {\it
Periods and Duality Symmetries in Calabi-Yau Compactifications},
preprint UCVFC-1-92.}.
It is a generalization of the $R\to 1/R$ symmetry of the bosonic
string compactified on $S^1$.

On the effective field theory level this string specific
symmetry is manifest insofar as the Lagrangian must be
invariant\ref\fslt{S. Ferrara, D. L\"ust, A. Shapere and S. Theisen,
\plb225(1989)363.}.
This has the important consequence that the moduli
dependent couplings must have definite transformation properties
under transformations of the duality group.
For the simplest case where the duality
group is just the modular group $SL(2,\Z)$, they are
modular forms. A possible non-perturbative potential for
the moduli fields must also respect this symmetry.

Let us illustrate this on a simple model. Since we are dealing with
string theories with $N=1$ space-time supersymmetry, the low-energy
effective action will be a K\"ahler sigma model. Consider the
case with one field only whose K\"ahler potential is
$K=-3\log(t+\bar t)$.
Let us assume (as is the case in simple orbifold compactifications)
that $t$ is the modulus field whose vacuum expectation value
determines the size of the six-dimensional compact space,
i.e. $t=R^2+ib$ where $R$ measures the size of the internal
manifold in units of $\sqrt{\alpha^\prime}$ and $b$, whose
presence is required by $N=1$ space-time supersymmetry
($t$ must be a chiral superfield) is the internal axion.
This vacuum expectation is however undetermined as there is
no potential for $t$ (it is a modulus).
The supergravity action, which is completely determined by the
K\"ahler potential, is invariant
under the continuous $SL(2;\R)$ isometries of the K\"ahler metric,
i.e.
under $t\to{at-ib\over ict+d}$ with $ad-bc\neq0$.
The invariance is broken by adding a (non-perturbatively
generated) superpotential $W(t)$ for the field $t$. The matter part
of the supergravity action is now described by a single real
function $G(t,\bar t)=K(t,\bar t)+\log W(t)+\log\bar W(t)$
\ref\cremmer{E. Cremmer, B. Julia, J. Scherk, L. Girardello
and P. van Nieuwenhuizen, \npb147(1979)105;
E. Cremmer, S. Ferrara, L. Girardello and A. Van Proeyen,
\npb212(1983)413.}.
Looking at the new terms in the action which arise from the addition
of the superpotential (e.g. the gravitino mass term), one finds
that the action is only invariant under those transformations
$t\to f(t)$ that leave $G(t,\bar t)$ invariant. It thus follows that
any non-trivial superpotential will break the continuous
$SL(2;\R)$ symmetry. This is all right as long as we can choose a
superpotential such that the action has a residual $SL(2;\Z)$
symmetry which reflects the stringy duality symmetry.
One thus needs that $G$ is a modular invariant functions. With
$K$ as given above, this entails that $W(t)$ transforms as
$W(t)\to e^{i\alpha}(ict+d)^{-3}W(t)$, where the phase may depend
on the (real) parameters of the transformation, but not on $t$.
We have thus found that the superpotential for the modulus field
must be a modular function of weight $-3$ (possibly with a
non-trivial multiplier system (the phase)). Such a function is
furnished by $\eta(t)^{-6}$, where $\eta(t)$ is the
Dedekind function. (This solution is not unique, since
one may always multiply by a function of the modular invariant
$j(t)$.)  One now takes the expression for $G$ and computes
the scalar potential. For the simple case described here, one finds
that it has a minimum in the fundamental region for $R\sim O(1)$,
i.e. the compactification is stable.
If we now add charged matter fields, we have to modify the
K\"ahler potential to include them.
Let us denote the charged matter field by $A$ and include it
in the K\"ahler potential in the following form:
$K=-\log\left\{(t+\bar t)^3-A\bar A(t+\bar t)\right\}$. (This is
the lowest order appearance of the charged matter fields in
the twisted sector of simple orbifold compactifications
\ref\dlk{L. Dixon, V. Kaplunovsky and J. Louis,
\npb329(1990)27.},
\ref\flt{S. Ferrara, D. L\"ust and S. Theisen,
\plb233(1989)147.}.)
For the K\"ahler potential to be invariant (up to a
K\"ahler transformation), we need to require the following
transformation properties for the  matter fields:
$A\to {A\over (ict+d)^2}$. Consequently, the Yukawa coupling,
which is the term in the superpotential cubic in $A$, must be,
up to a phase, a modular function of weight $+3$, e.g.
$\eta(t)^6$. (This is again not unique but
the arbitrariness may be fixed by going
to special points in moduli space where a simple formulation of the
underlying string theory, e.g. in terms of free fields, is valid
and the couplings can be computed.)
Its value at the minimum of the potential for
the field $t$ determines the strength of the Yukawa coupling.

The above program has been carried through for simple orbifold
compactifications only\flt. For general string compactifications
one does not know the duality group and in the few cases where
it has been determined, functions with definite weight are
generally not known.
Below we will discuss a way of determining the duality group
for simple Calabi-Yau compactifications from the monodromy of
the solutions to the corresponding Piccard-Fuchs equations, which
are the differential equations satisfied by the periods of the
Calabi-Yau manifold as functions of the moduli.

Above we have already mentioned the two different kinds of moduli
and that they appear as massless scalar fields with vanishing
potential in the low energy $N=1$ supersymmetric effective action.
Their K\"ahler metric is the Zamolodchikov metric on the space
of conformal field theories parametrized by the moduli.
It is given by the two-point function of the corresponding
truely marginal operators.
Using superconformal Ward identities it was shown in
\dlk~
that the moduli manifold has the direct
product structure ${\cal M}={\cal M}_{h_{1,1}}\times
{\cal M}_{h_{2,1}}$ where ${\cal M}_{h_{i,j}}$ are
K\"ahler manifolds with dimension $h_{i,j}$.
This result was first obtained in
\ref\seiberg{N. Seiberg, \npb303(1988)286.} using $N=2$ space-time
supersymmetry via the link between heterotic and
type II theories; i.e. that
the same (2,2) superconformal field theory
with central charge $(c,\bar c)=(9,9)$ could have been used
to compactify the type II rather than the heterotic string with
the former leading to $N=2$ space-time supersymmetry.
(Recall that for the heterotic string
the remainder of $(0,13)$ units of central
charge is used for the $E_8\times SO(10)$ gauge sector
where the $SO(10)$ factor combines with the $U(1)$ current
of the left moving $N=2$ SCA to $E_6$.)

What will be important in the following is the fact that
the moduli metric is blind as to which theory one
is compactifying and thus has to satisfy also
in the heterotic case the additional constraints which
come from the second space-time supersymmetry in type II
compactifications.

The constraints amount to the fact that in a special coordinate
system (called special gauge) the entire geometry of the
Calabi-Yau moduli space is encoded in two holomorphic
functions of the moduli fields, $\tilde {\cal F}_{(1,1)}$
and $\tilde {\cal F}_{(2,1)}$, where the subscript indicates that
there is one function for each type of moduli
\ref\sk{For reviews see
S. Ferrara, Mod. Phys. Lett. {\bf A6} (1991) 2175;
S. Ferrara and S. Theisen, in Proceeedings of the Hellenic Summer
School 1989, World Scientific; and references therein.}.
${\cal F}$
is called the prepotential in terms of which the K\"ahler
potential is given by
$$
K=-\ln \tilde Y\qquad{\rm with}\qquad \tilde Y=
i\left\(2(\tilde{\cal F}-\overline{\tilde{\cal F}})
-(\tilde{\cal F}_i+\overline{\tilde{\cal F}}_i)
(t^i-\bar t^{\bar\imath})\right\)\, ,\eqno{(1)}
$$
where $\tilde{\cal F}_i=\partial\tilde{\cal F}/\partial t^i$
and $t^i,\, i=1,\dots,h_{1,1},h_{2,1}$ are the moduli fields.
The Yukawa couplings are simply
$$
\kappa_{ijk}=-{\partial^3\over\partial t^i\partial t^j\partial t^k}
\tilde{\cal F}\,.
$$
The Riemann tensor on moduli space is then
$$
R_{i\bar\jmath k\bar l}=G_{i\bar\jmath}G_{k\bar l}
+G_{i\bar l}G_{k\bar\jmath}-e^{2K}\kappa_{ikm}
\kappa_{\bar\jmath\bar l\bar n}G^{m\bar n}\, ,
$$
where $G_{i\bar\jmath}={\partial^2\over\partial t^i
\partial t^{\bar\jmath}}K$ is the K\"ahler metric on moduli space.
One may introduce homogeneous coordinates on moduli space in terms
of which the prepotentials are homogeneous functions of
degree two (cf. below).
There is, of course, one set of above expressions for each
factor of moduli space corresponding to
$\tilde{\cal F}_{(1,1)}$ and $\tilde{\cal F}_{(2,1)}$.
K\"ahler manifolds with these properties are called
special.
Note that above expressions are not covariant and only
true in the special gauge. For the covariant formulation, see
\ref\astrom{A. Strominger, \cmp133(1990)163.},
\ref\sg{L. Castellani, R. D'Auria and S. Ferrara, \plb241(1990)57
and \cqg7(1990)1767; S. Ferrara and J. Louis, \plb278(1992)240.}.

As expected, these constraints on the K\"ahler structure are
inherited
from the Ward-identities of the underlying (2,2) super-conformal
algebra\dlk.

We have seen that the Yukawa couplings are given
by the third derivatives of the prepotentials with respect
to the moduli. This entails that they do not mix
the two sets of moduli and their corresponding matter fields;
i.e. the Yukawa couplings of the $\overline{27}'s$ of $E_6$ only
depend on the K\"ahler moduli and the couplings of the
$27's$ depend only on the complex structure moduli.
Whereas the former acquire contributions from
world-sheet instantons, the latter
do not\ref\digr{J. Distler and B. Greene, \npb309(1988)295.}
and are thus in general easier to compute.
In fact, the $27^3$ Yukawa couplings can be
evaluated exactly at the $\sigma$-model tree level or in
the point field theory limit. The absence of (perturbative
and non-perturbative) $\sigma$-model corrections is due
to the fact that the $\sigma$-model expansion parameter
$\alpha^\prime/R^2$ depends on one of the (1,1) moduli which,
as noted above, does not mix with the (2,1) moduli.

Let us now connect the above discussion with the cohomology
of the Calabi-Yau space $M$\ref\fs{S. Ferrara and A. Strominger,
in the Proceedings of the Texas A \& M Strings `89 Workshop;
ed. R. Arnowitt et al., World Scientific 1990.},
\ref\co{P. Candelas, P. Green and T. H\"ubsch \npb330(1990)49;
P. Candelas, X.C. de la Ossa,
\npb355(1991)455.}. Let $\alpha_a$ and $\beta^b$
($a,b=0,\dots,h_{2,1}$) be an integral
basis of generators of $H^3(M,\Z)$, dual to a canonical
homology basis $(A^a,B_b)$ for $H_3(M,\Z)$ with intersection
numbers $A^a\cdot A^b=B_a\cdot B_b=0,\, A^a\cdot B_b=\delta^a_b$.
Then
$$
\int_{A^b}\alpha_a=\int_M\alpha_a\wedge\beta^b=
-\int_{B_a}\beta^b=\delta_a^b
$$
with all other pairings vanishing. A complex structure on $M$
is now fixed by choosing a particular 3-form as the
holomorphic (3,0) form, which we will denote by $\Omega$.
It may be expanded in the above basis of $H^3(M,\Z)$ as
$\Omega=z^a\alpha_a-{\cal F}_a\beta^a$ where
$z^a=\int_{A^a}\Omega,\,{\cal F}_a=\int_{B_a}\Omega$
are called the periods of $\Omega$.
As shown in \ref\brgr{R. Bryant and P. Griffiths,
Progress in Mathematics {\bf 36}, p.77; Birkh\"auser, 1983.}
the $z^a$ are complex projective coordinates for the complex
structure
moduli space, i.e. we have ${\cal F}_a={\cal F}_a(z)$.
Considering now that under a change of complex structure
$\Omega$ changes as\ref\tian{G. Tian, in {\it Mathematical
Aspects of String Theory}, S.-T. Yau, editor, World Scientific 1987.}~
${\partial\Omega\over\partial z^a}=
\kappa_a\Omega+G_a$ where $G_a$ are (2,1) forms
and $\kappa_a$ is independent of the coordinates of $M$ it follows
that $\int\Omega\wedge{\partial\Omega\over\partial z^a}=0$.
Using the expression for $\Omega$ given above, we conclude that
${\cal F}_a={1\over 2}{\partial\over\partial z^a}(z^b{\cal F}_b)$,
or ${\cal F}_a={\partial{\cal F}\over\partial z^a}$ with
${\cal F}={1\over 2}z^a{\cal F}_a(z)$, ${\cal F}(\lambda z)
=\lambda^2 {\cal F}(z)$. The $(2,1)$ forms in the variation
of $\Omega$ also enter the expression for the metric on
moduli space which can  be shown to be
\ref\strom{A. Strominger, \prl55(1985)2547.}
$G_{a\bar b}=-\int G_a\wedge G_{\bar b}/\int\Omega\wedge\bar\Omega$
and can be written as $G_{a\bar b}=-\partial_a\partial_{\bar b}
\ln Y$ with
$Y=-i\int\Omega\wedge\bar\Omega=-i(z^a\overline{\cal F}_a
-z^{\bar a}{\cal F}_a)$.
If we now transform to inhomogeneous coordinates
$t^a=z^a/z^0=(1,t^i),i=1,\dots,h_{2,1}$
(in a patch where $z^0\neq 0$) we find
that ${\cal F}(z)=(z^0)^2\tilde{\cal F}(t)$ and
$Y=|z^0|^2\tilde Y$ with $\tilde Y$
as given in Eq.(1).
The Yukawa couplings are then $\kappa_{ijk}=-{\partial^3\over
\partial z^i\partial z^j\partial z^k}{\cal F}|_{z^0=1}=
\int\Omega\wedge{\partial^3\Omega\over\partial t^i
\partial t^j\partial t^k}|_{z^0=1}$. Since it follows from the
homogeneity of ${\cal F}$ that
$\int\Omega\wedge{\partial^2\Omega\over\partial z^a\partial z^b}=0$,
we find that under a change of coordinates $t^i\to\tilde t^i(t)$
the Yukawa couplings transform homogeneously.

In this discussion the choice of basis for $H_3(M,\Z)$ has not been
unique. In fact, any $\pmatrix{A^{\prime a}\cr B^{\prime}_b\cr}=
S\pmatrix{A^a\cr B_b\cr}$ with $S$ an integer matrix that leaves
$J=\pmatrix{\phantom{-}0&{\bf 1}\cr -{\bf 1}&0\cr}$
invariant\footnote{$^*$}{This means that $S^T J S=J=SJS^T$, i.e.
$S\in Sp(2h_{2,1}+2;\Z)$.}
(${\bf 1}={\bf 1}_{(h_{2,1}+1)\times(h_{2,1}+1)}$)
will lead to a canonical basis.
If we write $S$ in block form as $S=\pmatrix{a&b\cr c&d\cr}$ then
the
basis of $H^3(M,\Z)$ transforms as
$\pmatrix{\beta^\prime\cr \alpha^\prime\cr}=\pmatrix{c&d\cr a&b\cr}
\pmatrix{\beta\cr\alpha}$. Looking at the decomposition of
$\Omega=(z,\partial{\cal F})J\pmatrix{\beta\cr\alpha\cr}$ we
find that $\pmatrix{z^\prime\cr (\partial{\cal F})^\prime\cr}
=\pmatrix{c&d\cr a&b\cr}\pmatrix{z\cr\partial{\cal F}\cr}$.
Under these transformations $Y=-i(z,\partial{\cal F})J
\pmatrix{\bar z\cr\overline{\partial{\cal F}}\cr}$ is
also invariant, however not in general the prepotential
${\cal F}$.
Those $Sp(2h_{2,1}+2,\Z)$ transformations which act on the
homogeneous coordinates on moduli space as symmetries, i.e.
for which
${\cal F}^\prime={\cal F}$, are
referred to as duality transformations \ref\flt2{S. Ferrara,
D. L\"ust and S. Theisen, \plb242(1990)39.}.

So far we have only discussed the $(2,1)$ forms. An analogous
discussion for the (1,1) forms in term of a basis of $H^2(M,\Z)$
is also possible\fs,\co .
However as we have noted above, the point
field theory results obtainable in this way are only a small
part of the story, since they will get corrected
perturbatively and by instantons.
It is known \fs,\co~ that prior to  receiving quantum
corrections the prepotential
$\tilde{\cal F}_0$ for the (1,1) moduli space takes the form
$\tilde{\cal F}_0=-{1\over 6}\kappa_{ijk}t^i t^j t^k$,
where $t^i$ ($i=1,\dots, h_{1,1}$) are now the
(inhomogeneous) coordinates on
(1,1) moduli space and $\kappa_{ijk}$ are integral intersection
matrices of (1,1) forms $e_i$ which form a basis of
$H^2(M,\Z)$ and in terms of which we expand
$B+iJ=t^i e_i$ where $J$ is the K\"ahler form on $M$ and
$B$ the antisymmetric tensor field.
One can introduce homogeneous coordinates $\omega^a$ with
$t^a={\omega^a\over \omega^0}=(1,t^i),\,(a=0,\dots,h_{1,1})$
in terms of which ${\cal F}(\omega)=(\omega^0)^2\tilde{\cal F}$
is homogeneous of degree two.

Due to a perturbative non-renormalization theorem for the
Yukawa couplings\ref\witten{E. Witten, \npb268(1986)79} the
complete expression for the (1,1) prepotential must of the form
$\tilde{\cal F}=-{1\over 6}\kappa_{ijk}t^i t^j t^k
+{1\over 2}a_{ij}t^i t^j+b_i t^i +c +O(e^{-t})$ where the
polynomial part is perturbative and the non-polynomial part
due to instanton corrections which are, except for simple
torus compactifications, hopelessly difficult to compute directly. One
has to think of alternative ways to get at the full prepotential
(and thus the Yukawa couplings and the K\"ahler metric)
for the (1,1) sector.
This is where mirror symmetry, to be discussed next, enters
the stage.

On the conformal field theory level the $27's$
and $\overline{27}'s$ of $E_6$, and by world-sheet supersymmetry
the two types of moduli, can be simply interchanged by flipping
the relative sign of the left and right $U(1)$ charges of
the (2,2) superconformal algebra
\ref\grpl{B. Greene and Plesser,\npb338(1990)15.}.
On the geometrical level this corresponds to an interchange
of the Hodge numbers
$h_{1,1}$ and $h_{2,1}$ and thus to a change of sign of the
Euler number. This so called mirror map relates topologically
distinct Calabi-Yau spaces. The mirror hypothesis states that
the prepotentials for the different types of moduli are
interchanged on the manifold and its mirror.
Mirror symmetry thus allows
one to get the instanton corrected couplings for the
(1,1) forms on a given Calabi-Yau manifold $M$ from the
couplings of the (2,1) forms on its mirror $M^\prime$, which have no
instanton corrections.

The crucial question whether such a mirror manifold always
exists is not answered for general CY manifolds.
For special constructions, e.g. for the
(canonical desingularisations of) Fermat-type hypersurfaces
in weighted projective spaces $\IP(\us w)$ of dimension four,
it is known \ref\roan{S. S. Roan, Int. Jour. Math. {\bf 1}
(1990) 211; see also {\it Topological Couplings of Calabi-Yau
Orbifolds}, Preprint MPI f. Math. Bonn: MPI/92/22},
\ref\batyrev{V. V. Batyrev, {\it Dual polyhedra and the mirror
symmetry for Calabi-Yau hypersurfaces in toric varieties},
Preprint 92, Math. Fak. Univ. Essen} that a mirror manifold
$M^\prime$
with flipped Hodge diamond (i.e. $h_{i,j}\leftrightarrow h_{3-j,i}$)
can always be obtained
as (a canonical desingularisaton of) the orbifold of the Fermat
hypersurface
w.r.t. to its maximal abelian isotropy group $G_{max}$,
which acts locally as a subgroup of $SU(3)$.
The Fermat hypersurfaces are defined\footnote *{Underlined quantities
are five
tuples,
$\us x :=(x_0,\ldots,x_4)$ etc.}
 by
$M=X_k(\us w):=\{\us x\in
\IP(\us w)|W_0=\sum_{i=0}^4 a_i x_i^{n_i}=0\}$, $a_i\in \IC$ (we
set $a_i=k/n_i$ in the following), $n_i\in \N$.
The degree of $W_0$ is $k:={\rm lcm}\{\us n\}$ and
for the weights one chooses $w_i=k/n_i$ such that Eq. $W_0=0$ is
welldefined
on the equivalence classes $\lbrack \us x \rbrack $
of $\IP(\us w)$ (subject to $x_i\cong \lambda^{w_i} x_i$ with
$\lambda\in \IC\setminus \{0\}$). The map
$W_0:(\IC^5,\us 0)\rightarrow (\IC,0)$
is transversal in $\IP(\us w)$ as the only solution to $dW_0=0$ is
located
at $\us x=\us 0\notin \IP(\us w)$.
It is said to have an isolated singularity at the origin.
Nevertheless $X_k(\us w)$ can be singular as $W_0=0$ intersects in
general
the singular locus of $\IP(\us w)$. The latter one is described by
${\rm Sing}(\IP(\us w))=\bigcup_{I\subset
\{0,\ldots,4\}}\{\IP_I|c_I>1\}$,
where $\IP_I=\IP(\us w)\cap\{x_i=0,\forall i \in I\}$ and
$c_I:={\rm gcd}(w_j | j\in \{0,\ldots,4\},j\notin I)$.

Vanishing of the first Chern class $c_1=0$ requires\ref\yau{
S. T. Yau, {\it Compact three dimensional K\"ahler manifolds
with zero Ricci curvature}, in
{\it Geometry, Anomalies, Topology (1985)},
Eds. W. Bardeen, A. White}
$$
\sum_{i=0}^4 w_i=k\,;\eqno{(2)}
$$
it renders the number of $X_k(\us w)'s$ finite. Eq. $(2)$ implies
also
that $X_k(\us w)$ has only singular points and singular curves. Due
to Eq. (2)
the corresponding singularities are moreover of
Gorenstein-type\roan,\batyrev  and
can be resolved in a canonical way to a Calabi-Yau manifold.
The resolution process introduces new elements in the Hodge
cohomology
$H^{1,1}$ (and $H^{2,1}$). The only examples for which this does not
occur,
because $X_k(\us w) \cap {\rm Sing} (\IP(\us w))=\emptyset$
are: $X_5(1,1,1,1,1)$, $X_6(2,1,1,1,1)$, $X_8(4,1,1,1,1)$ and
$X_{10}(5,2,1,1,1)$. Here the only class of form degree (1,1) is the
pullback of the K\"ahler class of $\IP(\us w)$.

There are strong indications that string theory on Fermat
CY manifolds --- at a special point of the moduli space ---
correspond to string compactifications on special Gepner type
models\ref\gep{D. Gepner, {\it The Exact Structure of Calabi-Yau
Compactification}, in {\it Perspectives of String Theory} Copenhagen
1987, Ed. P. Di Vecchia, J. L. Peterson}.
These are tensor products of five (four) $n=2$ superconformal
$SU(2)/U(1)$ coset models (minimal $n=2$ series),
where the left and right characters are tied together according to
the $A$-type affine modular invariant, i.e.
diagonally\footnote{$^{**}$}{The power $n_i$ is related to the level
$p_i$
of the $i'$th
minimal factor model  by $n_i=p_i+2$. At most one $n_i$ can be 2, in
this case one has only
four nontrivial factor models.}. The correpondence is established
at the level of cohomology, i.e. the dimensions of the cohomology
groups
and parts of the ring structure in cohomology on the Calabi-Yau
manifold coincide with the one of the cohomology of the
$(chiral,chiral)$ and $(chiral,antichiral)$
rings\ref\lvw{W. Lerche, C. Vafa, N. P. Warner \npb324(1989)427} in
the $n=2$
superconformal theory.

Let us look at the correspondence between the two constructions
at the level of the discrete symmetries.
Each of the factor theories has a $(Z_{p+2}\times Z_2)$
symmetry\footnote{$^\dagger$}{There exist left
and right versions of these symmetry.
We restrict ourselves here to the left-right symmetric subgroup.
Permutation symmetries, which are present whenever several tensor
theories are identically carry trivially over to the manifold.}.
The partition function of the heterotic string theory is constructed
by
orbifoldisation of the internal tensor theory together with the
external contributions  w.r.t. a subgroup of these symmetries namely
$G_0=Z_{lcm\{p_i+2\}}\times Z_2^5$ and contains as the residual
invariance
${\cal G}=\prod_{i=1}^5 Z_{p_i+2}/Z_{lcm\{k_i+2\}}$.
The latter is in one to one correspondence with the discrete symmetry
group on the hypersurface $X_k(\us w)$ generated by $x_i\mapsto
\exp \lbrack 2 \pi i a_i/n_i\rbrack x_i$. Note that $Z_{lcm\{k_i+2\}}$ acts
trivially on the equivalence classes $[\us x]$ of the $\IP(\us w)$.
One can construct new heterotic string theories by dividing out
subgroups of
$\cal G$, which leave the space-time supersymmetry operator, a
conformal
field in the Gepner model, invariant\ref\fkss{J. Fuchs, A. Klemm,
C. Scheich, M. Schmidt, \plb232(1989)383, Ann. Phys. {\bf 204} (1990)
163}.
This is the case if
$$
\sum_{i=1}^5 a_i/n_i\in \IZ\eqno{(3)}
$$
$\forall $ their generators and is analogeous to the geometrical
requirement
that the group acts trivially on the holomorphic $(3,0)$-form.
In the formalism of orbifold constructions of CFT it is possible
to prove\gep,\ref\ak{A. Klemm, Ph.D. Thesis, Heidelberg University,
Preprint HD-THEP-90-45} that these new models appear in mirror
pairs and that
the mirror of a model is obtained by orbifoldising w.r.t.
the maximal subgroup $G_{max}$ of $\cal G$ subject to condition
$(3)$.
Moreover the only difference in the partion function of these
mirror pairs is a sign flip of the $U(1)$ charge of the
holomorphic sector relative to the antiholomorphic sector.

The procedure of dividing out these subgroups $G_i$ of $\cal G$
can also be performed on the hypersurface $X_d(\us w)$. The
orbit space $X_k(\us w)/G_i$ is in general singular due to fixed
point
singularities. With help of $(2)$ and $(3)$
one can show that the singular locus consists again only of
points and curves. Furthermore all the singularities are of
Gorenstein-type and can be desingularized canonically
to a smooth Calabi Yau manifold.
The string theories described by the geometric desingularisation of  the
orbit space
and the CFT orbifold coincide at the same level as the original theories
do,
namely in parts of their cohomology structure and their symmetries
\fkss,\ak. As mentioned above the mirror $M^\prime$ is given by the
canonical desingularisation $M_k^\prime=\widehat{X_k(\us w)/G_{max}}$
\roan.

There exists an elegant and mathematically rigorous formulation of
the occurence of mirror symmetry in the context of toric varieties,
which includes all orbifolds of the Fermat-type hypersurfaces
mentioned above.
The data of the space are encoded in a pair of
reflexive polyhedra with integral vertices and a lattice.
The Fermat-type hypersurface and their orbifolds
can be constructed from  pairs of simplicial, reflexive polyhedra and
a lattice by means of toric geometry.
It is shown in \batyrev that the same construction applied to the
dual polytope in the dual lattice gives rise to the mirror
configuration likewise represented as a hypersurface in a toric
variety.

The mirror hypothesis implies a one to one map between the
moduli space of the complex structure moduli on the
manifold and the K\"ahler structure moduli of its mirror.
The close relation to the CFT theories and properties of their
orbifolds mentioned above suggest that such a map exists, at least
locally, in the vicinity of the exactly solvable pair and can therefore
be
extended
-- possibly  not uniquely -- to the whole moduli space.

As we have seen, the physically relevant quantities, namely the
K\"ahler potential and the Yukawa couplings for the sector
of the theory which depends on to the complex structure moduli,
can be calculate from the period functions.
If the mirror hypothesis
is correct one can obtain the same information for the
sector which depends on the K\"ahler moduli from the periods
of the mirror manifold\cdgp,\kt,\afont.
The periods are known to satisfy linear differential
equations, called Piccard-Fuchs equations.
To illustrate this we consider the torus $T^2$ defined as the
algebraic curve $y^2=x(x-1)(x-\lambda)$.
Consider the differential $\Omega(\lambda)={dx\over y}$
whose integrals over the two non-trivial homology cycles
are the periods.
Since the first Betti number $b_1(T^2)=2$ there must exist a
relation between the three differentials
$\Omega,\,{\partial\Omega\over\partial\lambda}$ and
${\partial^2\Omega\over\partial\lambda^2}$. Some linear combination
with coefficients being functions of $\lambda$ must be an excact
differential whose integral vanishes upon integration
over a closed cycle; i.e. the periods of the torus satisfy
a linear ordinary second order differential
equation\footnote{$^*$}{It might be interesting
to note that the differential equations one gets from
the requirement of the vanishing of the curvature of the
metric in coupling constant space
\ref\dvv{R. Dijkgraaf, H. Verlinde and E. Verlinde,
\npb352(1991)59; A. Lossev, lectures at this conference.}.
for the three $c=3$
topological Landau-Ginzburg theories are exactly the
Picard-Fuchs equations for the tori these theories are
orbifolds of\ref\kts{A. Klemm, S. Theisen and M.G. Schmidt,
{\it Correlation Functions for Topological Landau-Ginzburg
Models with $c\leq 3$}, \ijmpa (in press).}.}.
We will denote the periods by
$\Pi_i=\int_{C_i}\omega$.

The generalization to more complicated cases, including
higher dimensional manifolds and more than one modulus
is straightforward, in the latter case leading to systems of
partial differential equations. In the following we will restrict
ourselves to the case of one modulus only and consider the
Fermat CY manifolds (see above). Here the periods are defined
as above, namely as the integrals of the holomorphic three
form over the $H^3(M,\Z)$ cycles. Since $b_3=\sum_{p+q=3}h_{p,q}=4$,
the differential equation satisfied by the periods as functions
of the one complex structure modulus,
which we will denote by $\alpha$,
will be of fourth order
whose four solutions
correspond to the four periods. The problem will be to find the
correct
linear combinations of the solutions such that they correspond
to the periods of $\Omega$ expanded in the basis of integer
cohomology dual to the canonical cycles $(A^a,B_b)$.

The Calabi-Yau spaces that so far have been amenable to above
treatment are the ones mentioned before, which have only one K\"ahler
modulus
(see \cdgp for the case $k=5$ and \kt,\afont
for all four cases (see also \ref\morrison{ D. R. Morrison,
{\it Picard-Fuchs equations and mirror maps for hypersurfaces},
Preprint DUK-M-91-14.})).
Allowing for all possible deformations of the complex
structure they take the form
$$
X_k(\us w)=\left\{ x_i\in\IP(\us{w})|
W\equiv W_0-\sum a_{ijklm}x_0^i x_1^j
x_2^k x_3^l x_4^m=0\right\}
$$
with $W_0$ as given above. The deformations of $W_0$
are the elements in the polynomial ring
${\cal R}={C\(x_i\)\over dW_0}$ with the same degree as $W_0$.
The coefficients $a_{ijklm}$ parametrize ${\cal M}_{(2,1)}$ and
one finds $h_{2,1}=101,103,149,145$ for $k=5,6,8,10$
respectively, corresponding to Euler numbers
$\chi=2(h_{1,1}-h_{2,1})=-200,-204,-296 ,-288$. Their mirrors are
obtained by dividing by the full phase symmetry  group $G_{max}$ which is
$\Z_5^3,~\Z_3\times\Z_6^2,~\Z_2\times\Z_8^2$ and $\Z_{10}^2$
for the four cases considered. The only surviving deformation is
then $\alpha\equiv a_{11111}$ and ${\cal R}$ consists of the
elements $(x_0\dots x_4)^\lambda,\,\lambda=0,1,2,3$ only.
Indeed, by restricting to this invariant subring, we essentially
study the complex structure deformation of the
mirror manifold, which has $h_{2,1}=1$.
One may verify the interchange
of the Hodge numbers $h_{2,1}$ and $h_{1,1}$ by explicit
construction of the geometric desingularisation. With a suitable
choice of constants in $W_0$ (namely $a_i={k\over n_i}$) and
$\alpha\to k\alpha$, $\alpha^k=1$ are nodes of the four manifolds.
They become singular at $\alpha\to\infty$.

To set up the Picard-Fuchs equations
\ref\griff{P. Griffiths, Ann. Math. {\bf 90} (1969) 460;
see also W. Lerche, D. Smit and N. Warner, \npb372(1992)87.},
\ref\cf{A. Cadavid and S. Ferrara, \plb267(1991)193.},\kts,
\ref\louis{J. Louis, {\it Differential Equations in Special K\"ahler
Geometry}, preprint CERN-TH. 6580/92.}
we need an explicit expression
for the periods\griff,\ref\can{P. Candelas, \npb298(1988)458.}.
If $\gamma$ is a small circle winding around the
hypersurface $W=0$, we may represent the holomorphic three form
$\Omega$ as
$$
\int_\gamma{q(\alpha)\over W(\alpha)}\omega\qquad{\rm with}\qquad
\omega=\sum_{i=0}^4(-1)^i x^i dx^0\wedge\dots\wedge
\widehat{dx^i}\wedge\dots\wedge dx^4
$$
where the hat denotes omission. Obviously, under
$x^i\to\lambda x^i$, $\Omega$ does not change and is thus
a (nowhere vanishing) three form on $\IP_4$.
The function $q(\alpha)$ reflects the gauge freedom of
$\Omega$, which is a holomorphic section of the projective
line bundle associated to the Hodge bundle over ${\cal M}_{(2,1)}$
with fibers $H^3(M)$ \astrom.
The periods are then
$\Pi_a=\int_{\Gamma_a}{q(\alpha)\over W(\alpha)}\omega$,
where $\Gamma_a$ is a 4-cycle in $\IP_d-M$ which is homologous
to a tube over a three-cycle on $M$. This is shown in \griff
where one also finds a proof of the fact that one may
integrate by parts with respect to the coordinates of $\IP_4$.

For the purpose
of deriving the period equation, it is most convenient to
set $q(\a)=1$. Differentiating $\lambda$ times with respect to $\a$
produces terms of the form $\int{(x_0 x_1 x_2 x_3 x_4)^\lambda
\over W^{\lambda+1}(\a)}\o$. The $\lambda=4$ term, which is the first
to produce an integrand whose numerator is no longer in the
ring ${\cal R}$, can be expressed, using the expressions
$\partial W/\partial x_i$
and integration by parts, in terms of lower derivatives.
The computation is straightforward and
produces
$$
\eqalign{
k=5:\,\,&(1-\a^5)\,\Pi^{(iv)}-10\,\a^4\, \Pi^{\prime\prime\prime}\,
-25\,\a^3\, \Pi^{\prime\prime}-15\,\a^2\, \Pi^{\prime}\,- \a\,\Pi=0\cr
k=6:\,\,
&\a^2(1-\a^{6})\,\Pi^{(iv)}-2\a(1+5\a^6)\,\Pi^{\prime\prime\prime}\,
+(2-25\a^6)\,\Pi^{\prime\prime}-15\a^5\,\Pi^\prime-\a^4\,\Pi=0\cr
k=8:\,\,
&\a^3(1-\a^8)\Pi^{(iv)}-\a^2(6+10\a^8)\Pi^{\prime\prime\prime}
+5\a(3-5\a^8)\Pi^{\prime\prime}-15(1+\a^8)\Pi^\prime-\a^7\Pi=0\cr
k=10:\,\,&\a^3(1-\a^{10})\,\Pi^{(iv)}
-10\a^2(1+\a^{10})\,\Pi^{\prime\prime\prime}\cr
&\quad
+5\a(7-5\a^{10})\,\Pi^{\prime\prime}-5(7+3\a^{10})\,\Pi^\prime-\a^9\,\Pi=0}
$$
A fundamental system of solutions may be obtained
following the method of Frobenius for ordinary differential
equations with regular singular points \ref\ince{E.L. Ince,
{\it Ordinary differential equations}, Dover 1956.}
which are here
$\a=0$, $\a=\infty$ and $\a^k=1$.
The solutions of the indicial equations
at the three singular points are
$\rho=(0,1,2,3)_{k=5},~(0,1,3,4)_{k=6},~(0,2,4,6)_{k=8},~
(0,2,6,8)_{k=10}$ for $\a=0$, $\rho=(0,1_2,2)$ for $\a^k=1$
and $\rho=0_4$ for $\a=\infty$. The subscripts denote the
multiplicities of the solutions. It follows from the
general theory that at $\a=\infty$ there is one solution
given as a pure power series and three containing logarithms
(with powers 1,2 and 3, respectively). At $\a=0$, all four
solutions are pure power series as one sees e.g. by
noting that we can rewrite the differential equation
in terms of the variables $\a^k$, for which the solutions of
the indicial equation would no longer differ by integers.
The point $\a=1$ needs some care. (The other solutions
of $\a^k=1$ are treated similarly). There is one power series
solution with index $\rho=2$ and at least one logarithmic
solution for $\rho=1$.
Making a power series ansatz for $\rho=0$ one finds that the
first three coefficients are arbitrary which means that there is
one power series solution for each $\rho$. One also easily
checks that in the second solution to $\rho=1$ the
logarithm is multiplied by a linear combination of the power
series solutions with indices 1 and 2.
To summarize, the periods of the manifolds have logarithmic
singularities at the values of $\a$ corresponding to the
node ($\a^k=1$) and to the singular manifold ($\a=\infty$).
We will thus get non-trivial monodromy about these points.

With reference to the literature \cdgp,\kt,\afont
we will skip the details of the computation which is a
sophisticated exercise in the theory of linear ordinary
differential equations with regular singular points.

As we have discussed above, in order to get the prepotential
from the solutions of the period equation we have to find a basis
in which the monodromy acts as $SP(4,\Z)$ transformations and
${\cal F}$ is then given as ${\cal F}={1\over 2}{\cal F}_a z^a$.
This can be achived since it is possible to compute two of the
periods explicitly. Then, up to a $SP(2,\Z)\subset SP(4,\Z)$
transformation which acts on the remaining two periods,
this basis can be found.

Again skipping details \cdgp,\kt, we simply give the results for
various
quantities of interest in the limits of large and small
values of the modulus $\alpha$. For the K\"ahler potential
and K\"ahler metric we find
$(\gamma=k\prod_{i=0}^4(w_i)^{-w_i/k})$

\noindent $\a\to\infty$:
$$
e^{-K}\simeq{(2\pi)^3\over{\rm Ord}\,G}
\left({4k\over 3}\log^3|\g\a|+{2\over 3k^2}
\left(k^3-\sum_{i=0}^4 w_i^3\right)\zeta(3)\right)
$$
$$
g_{\a\bar\a}\simeq{3\over 4|\a|^2\log^2|\g\a|}
\left(1+{2\left({\displaystyle\sum_{i=0}^4}\left({w_i\over
k}\right)^3
-1\right)\zeta(3)\over\log^3|\g\a|}\right).
$$
In terms of the variable $t\propto i\log(\g\a)$
the leading behaviour is $g_{t\bar t}\simeq-{3\over(t-\bar t)^2}$
which is the metric for the upper half plane with curvature
$R=-{4/ 3}$.
\hfil\break
\noindent $\a\to 0$:
$$
\mat{&e^{-K}_{k=5}=\displaystyle{{(2\pi)^3\over
5^5}{\Gamma^5({1\over5})\over
                   \Gamma^5({4\over 5})}\,|\a|^2\,+\,
                   O(|\a|^{4})};\quad&
     &e^{-K}_{k=6}=\displaystyle{{2^{13/3}\pi^8\over 3^{11/2}
                   \Gamma^2({2\over 3})
                  \Gamma^8({5\over 6})}\,|\a|^2\,+\,O(|\a|^{4})},\cr
\noalign{\medskip}
     &e^{-K}_{k=8}=\displaystyle{{\pi^7\over 128}{\cot^2({\pi\over
8})
                   \over\Gamma^8({7\over 8})}\,|\a|^2
                   \,+\,O(|\a|^{6})},&
     &e^{-K}_{k=10}\simeq 104.61\,|\a|^2\,+\,O(|\a|^{6});\cr}
$$
$$
\mat{
         &{g}^{k=5}_{\a \bar \a}=
         \displaystyle{ 25 \left(\Gamma({4\over 5}) \Gamma({2\over
5})\over
         \Gamma^3({1\over 5}) \Gamma({3\over
5})\right)^5+O(|\a|^2)},\qquad
         &{g}^{k=6}_{\a \bar \a}=\displaystyle{
         {3 \Gamma^8({5\over 6})\over
         {{2^{{2\over 3}}}\,{{\pi }^2}} \Gamma^4({2\over 3})}\,+
O(|\a|^2)},
         \qquad \cr
\noalign{\medskip}
         &{g}^{k=8}_{\a \bar \a}=\displaystyle{
         {64 (3-2^{3/2})^2 \Gamma^8({7\over 8})\over
          \Gamma^8({5\over 8})}\,\, |\a|^2\, + O(|\a|^8)},\,\,\qquad
          &{g}^{k=10}_{\a \bar \a}\simeq
          0.170 \,\, |\a|^2\, +\,O(|\a|^6).\qquad }
$$

The invariant Yukawa couplings are defined as
$$
{\cal Y}_{inv}=g_{\a\bar\a}^{-3/2} e^K
|\kappa_{\a\a\a}|
$$
where $\kappa_{\a\a\a}=\int\Omega\wedge{\partial^3\Omega
\over\partial\a^3}$.
They correspond to a canonically normalized kinetic energy
of the matter fields (hence the factor $g_{\a\bar\a}^{-3/2}$)
and are invariant under K\"ahler gauge
transformations induced by moduli-dependent rescalings of $\Omega$
(hence the factor $e^K$). For the cases under
consideration we found
$\kappa_{\a\a\a}=(2\pi i )^3 k \alpha^{k-3}/({\rm Ord
G}\,(1-\alpha^k))$.

In the limits considered above we find for the
leading terms
of the Yukawa couplings of the one multiplet of
$27$ of $E_6$:
\hfil\break\noindent$\a\to \infty$:
$$
{\cal Y}_{inv}={2\over\sqrt{3}}\,\quad\forall\,k\,.
$$
\noindent$\a\to 0$:
$$
   \mat{
         &{\cal Y}^{k=5}_{inv}=
         \displaystyle{\left(\Gamma^3({3\over 5}) \Gamma({1\over
5})\over
         \Gamma^3({2\over 5}) \Gamma({4\over 5})\right)^{5\over 2}+
          O(|\a|^2)},\qquad
         &{\cal Y}^{k=6}_{inv}=\displaystyle{2^{4\over 3}}\,\,
         |\a|  \,+ O(|\a|^3),
         \qquad \cr
\noalign{\medskip}
         &{\cal Y}^{k=8}_{inv}=\displaystyle{
        {\Gamma^6({5\over 8}) \Gamma^2({1\over 8})\over
         \Gamma^6({3\over 8}) \Gamma^2({7\over 8})}+O(|\a|^2)}
,\,\,\,\,\,\,
          &{\cal Y}^{k=10}_{inv}=
          3.394 \,\, |\a|^2\, +\,O(|\a|^6)\,.}
$$

For $k=5,8$ the nonvanishing couplings coincide with the values
of the corresponding Gepner models, which can be calculated using
the relation\digr
between the operator product coefficients of the
minimal $(n=2)$ superconformal models and the known ones of the
$su(2)$ Wess-Zumino-Witten theories. In the $k=6,10$ cases the
additional $U(1)$ selection rules at the Gepner point exclude the
coupling,
which is allowed for generic values of the modulus.

This closes the first part of the program. We have found the
exact (due to absence of $\sigma$-model corrections)
K\"ahler potential and Yukawa couplings for the (2,1) sector
of the moduli space of the CY spaces $M_k$.

To get the couplings for the single (1,1) form of the
original manifold, one has to perform the mirror map. This
way we will obtain the complete expression, i.e. including
all (instanton) corrections, e.g. for the Yukawa couplings. This
then contains also information about the numbers on instantons
(rational curves) on the original manifold, information
otherwise hard to obtain\cdgp,\kt,\afont,\morrison.

As already mentioned,
the (1,1) sector of the original manifold is also described by a
holomorphic function
${\cal F}$
which is homogeneous of degree two.
The large radius limit of ${\cal F}$ is known; it takes
the simple form ${\cal F}_0=-{\kappa_0\over 6}
{(\omega^1)^3\over\omega^0}=-{\kappa_0\over 6}(\omega^0)^2 t^3
=(w^0)^2\tilde{\cal F}_0$
where $t={\omega^1\over\omega^0}$ is the inhomogeneous coordinate
of the (1,1) moduli space. $\kappa_0=-\partial_t^3\tilde{\cal F}_0$
is the infinite radius limit of the Yukawa coupling
and is given by an intersection number. The latter evaluate
to  $\kappa_0=\{5,3,2,1\}$ for $k=\{5,6,8,10\}$ for the
manifolds under consideration \yau ,\kt.
Like the Yukawa coupling(s) the K\"ahler potential
derives from $ \tilde {\cal F} $ as in Eq. (1).
One finds $(t=t_1+it_2)$
$$
K_0=-\log\left({4\kappa\over 3}t_2^3\right)\,.
$$
{}From this we easily arrive at the large
radius limits of the metric $g^0_{t\bar t}={3\over 4t_2^2}$
and of the Ricci tensor $R^0_{t\bar t}=-{2\over 3}g^0_{t\bar t}$.
For the Ricci scalar one thus gets $R^0=-{4\over 3}$ and
for the invariant Yukawa coupling ${\cal Y}_0={2\over\sqrt{3}}$.
These same constant values were found as the large complex structure
limits for the (2,1) moduli spaces of the mirrors $M_k^\prime$.

As discussed before,
these infinite radius results get modified by sigma model
loops and instanton contributions, the latter being
non-perturbative in the sigma model expansion parameter
$1/R^2\sim 1/t$, $R$ being a measure for the size of the manifold.
The fully corrected prepotential
has the form
$$
\tilde{\cal F}=-{\kappa_0\over6}t^3
+{1\over 2}at^2+bt+c+O\Bigl(e^{-t}\Bigr).\eqno(4)
$$
The polynomial part is perturbative and restricted
by the perturbative
non-renormalization theorem for Yukawa couplings;
note that only imaginary parts of
$a,b$ and $c$ do affect the K\"ahler metric.

The mirror hypothesis implies now that the two prepotentials
for the $(2,1)$ modulus on the mirror and the $(1,1)$ modulus
on  $X_k(\us w)$ are essentially the same, but generally expressed
in two different symplectic bases for the corresponding
period\footnote{$^*$}{Of course we can define and calculate
the periods as integrals over cycles only on the mirror. The
`period' vector depending on the $(1,1)$ modulus is derived
from the corresponding prepotential Eq.(4) and has
components $(\omega^0,\omega^1,\partial{\cal F}/\partial\omega^0,
\partial{\cal F}/\partial\omega^1)$.}
vectors.
We have already seen that in terms of the variable $t\propto
i\log(\g\a)$ the large complex structure and large
radius limits of the K\"ahler metrics for the moduli spaces of the
(2,1) and (1,1) moduli agree.
By comparing the large radius limit with the large complex
structure limit one also determines an integer symplectic matrix
which
relates the period vectors up to a gauge transformation
which expresses the freedom in the definition of $\Omega$,
i.e. the fact that it is a section of a projective bundle.
This also fixes the coefficients $a,b,c$ in eq.(4).
$a$ and $b$ turn out to be real and the quadratic and linear term
do thus not contribute to the K\"ahler potential.
$c$ on the other hand is imaginary $\propto\zeta(3)$ and has
been identified in \cdgp\  with the
four loop contribution calculated in \ref\gvz{M. Grisaru,
A. van de Ven and D. Zanon, \plb173(1986)423, \npb277(1986)388,
\npb277(1986)409.}. This term also makes its appearance in
the effective low-energy string actions extracted from tree level
string scattering amplitudes\ref\grsl{D. Gross and J. Sloan,
\npb291(1987)41; N. Cai and C. Nunez, \npb287(1987)279.}.

The relation between $t$ and $\a$ is
($\Phi(N)={1\over k}\sum_{i=0}^4 w_i\psi(1+w_i N)-\psi(1+kN),\,
\psi(x)=d\log\Gamma(x)/dx$)
\def\o{\omega}
$$
t={\o^1\over\o^0}=-{k\over 2\pi
i}\left\lbrace\log(\g\a)
+{{\displaystyle \sum_{N=0}^\infty}{(kN)!\over\prod_{i=0}^4(w_i N)!}
\phi(N)(\g\a)^{-kN}\over
{\displaystyle \sum_{N=0}^\infty}{(kN)!\over\prod_{i=0}^4(w_i N)!}
(\g\a)^{-kN}}\right\rbrace\eqno{(5)}
$$
where the second expression is valid for $\a$ large.
Using the monodromy matrices for the periods on the
mirror one finds that as $\alpha$ is carried around infinity,
$t\to t+k$.

To get the Yukawa coupling we transform $\kappa_{\a\a\a}$
to the coordinate $t$ and find that the infinite radius
value $\kappa_0$ gets corrected to
$$
\kappa_{ttt}=\left({\omega^0\over{\cal G}_2}\right)^2
\kappa_{\a\a\a}\left({d\a\over dt}\right)^3\,.
$$
The prefactor expresses the gauge freedom and is due to the
relative factor (besides the integer symplectic matrix)
we have chosen between the two `period vectors'.
Its components
appear in the definition of the holomorphic three form which
enters quadratically in $\kappa_{\a\a\a}$.
In the gauge $\omega^0=1$ this becomes $\kappa_0+O(q)$ with
$q=\exp(2\pi i t)$,
where the instanton contributions come with integer
coefficients.
Indeed, on inverting the series (5)
and expressing the result in the form
$\kappa_{ttt}=\kappa_0+\sum_{j=1}^\infty{n_j j^3 q^j\over 1-q^j}$
conjectured in \cdgp\  and proven in \ref\asmo{P.S. Aspinwall
and D.R. Morrison, {\it Topological field theory and
rational curves}, preprint OUTP-91-32P.} we find the numbers
$n_j$ which count the rational curves of degree $j$ in $M$
\cdgp,\kt,\afont.

One can now also study
the duality symmetry of those models. The details can be
found in refs.\cdgp,\kt,\afont.
The Yukawa coupling will have a simple transformation
law under duality transformations. This follows from the fact
that the one matter superfield which is related to the
modulus via world-sheet supersymmetry will transform
homogeneously and to have an invariant  supergravity action
the Yukawa coupling must also transform
homogeneously\flt .
Having computed the Yukawa couplings we thus have an explicit
function of the modulus, which, when raised to the
appropriate power, is also a candidate for a non-perturbative
superpotential for the modulus itself.
Of course, whereas for the modular group $SL(2;\Z)$ this function
is known to be more or less unique, practically nothing is
known about automorphic functions of the groups one encounters
here.

The models considered represent only a very restricted
class. To make further progress towards realistic models
one has to extend the analysis in several directions.
One is to consider models described by
higher dimensional projective varieties. There are a few
examples of this kind with $h_{2,1}=1$, which can be studied
as a first step in this direction. Another generalization is
to models defined by more than one polynomial constraint.
The other obvious direction
to go is to consider models with more than one modulus,
leading to partial differential equations for the periods.
This seems to be the hardest of the possible generalizations.
\nobreak

\vskip.5cm
\noindent Acknowledgements: One of us (S.T.) would like to take
this opportunity to thank the organizers of the conference
for the invitation and the hospitality, and all participants
for the wonderful time in Kiev.
\vskip1cm

\footatend%\vfill\eject
\immediate\closeout\rfile\writestoppt
\baselineskip=14pt\centerline{{\bf References}}\bigskip{\frenchspacing%
\parindent=20pt\escapechar=` \input refs.tmp\vfill\eject}\nonfrenchspacing

\end